\documentclass[aps,prb,twocolumn,showpacs]{revtex4}

\usepackage{graphicx}
\usepackage{dcolumn}
\usepackage{amsmath}
\usepackage{amssymb}

\begin{document}

\markboth{Yu Lan, Xixiao Ma, Ling Qin, Yongjun Wang and Shiping Feng}{Asymmetric doping dependence of superconductivity between hole- and electron-doped triangular-lattice superconductors}

\title{Asymmetric doping dependence of superconductivity between hole- and electron-doped triangular-lattice superconductors}

\author{Yu Lan$^{1,2}$\footnote{Corresponding author. E-mail: ylan@bnu.edu.cn}, Xixiao Ma$^{3}$, Ling Qin$^{4}$, Yongjun Wang$^{5}$, and Shiping Feng$^{5}$\footnote{Corresponding author. E-mail: spfeng@bnu.edu.cn}}

\address{$^{1}$Hunan Provincial Key Laboratory of Intelligent Information Processing and Application, College of Physics and Electronic Engineering, Hengyang Normal University, Hengyang 421002, China}

\address{$^{2}$Department of Physics, The University of Texas at Dallas, Richardson, Texas 75080-3021, USA}

\address{$^{3}$The Patent Office of the State Intellectual Property Office, Beijing 100088, China}

\address{$^{4}$College of Physics and Engineering, Chengdu Normal University, Chengdu 611130, China}

\address{$^5$Department of Physics, Beijing Normal University, Beijing 100875, China}

\begin{abstract}
Within the framework of kinetic-energy-driven superconductivity, the asymmetric doping dependence of superconductivity between the hole- and electron-doped triangular-lattice superconductors has been studied. It is shown that although the superconducting transition temperature has a dome-shaped doping dependence for both the hole- and electron-doped triangular-lattice superconductors, superconductivity appears over a wide doping of range in the hole-doped case, while it only exists in a narrow range of the doping in the electron-doped side. Moreover, the maximum superconducting transition temperature around the optimal doping in the electron-doped triangular-lattice superconductors is lower than that of the hole-doped counterparts. The theory also shows that the asymmetric doping dependence of superconductivity between the hole- and electron-doped cases may be a common feature for a doped Mott insulator.
\end{abstract}

\vskip 0.5cm

\vskip 0.5cm

\maketitle

The study of oxide compounds has been of theoretical and experimental interest for a long time, since most of these oxide compounds are belong to the strongly correlated electron systems and exhibit unconventional superconducting (SC) mechanism \cite{Aoki04,Vojta17}, including the cuprate superconductors \cite{Bednorz86,Tokura89,Horio16,Song17} and cobaltate superconductors \cite{Takada03,Schaak03,Milne04,Sakurai06,Michioka06}. In particular, the parent compounds of the cuprate superconductors are a form of non-conductor called a Mott insulator, where the one-half spin Cu$^{2+}$ ions in a {\it square} array are ordered antiferromagnetically \cite{Fujita12}. When this parent square-lattice Mott insulator is doped with holes or electrons, superconductivity emerges \cite{Bednorz86,Tokura89,Horio16,Song17}, where the SC transition temperature $T_{\rm c}$ exhibits an unusual dome-shaped doping dependence. Moreover, although the strong electron correlation is common for both hole- and electron-doped square-lattice cuprate superconductors, the maximum achievable $T_{\rm c}$ in the electron-doped case is lower than that in the hole-doped side \cite{Bednorz86,Tokura89,Horio16,Song17}. On the other hand, the one-half spin Co$^{4+}$ ions in the undoped cobaltates sites sit on a {\it triangular-planar lattice}. The cobaltate superconductor Na$_{x}$CoO$_{2}\cdot y$H$_{2}$O is viewed as an electron-doped triangular-lattice superconductor \cite{Takada03,Schaak03,Milne04,Sakurai06,Michioka06}, since as electrons are introduced into these undoped cobaltates, superconductivity appears with $T_{\rm c}$ that has the same dome-shaped dependence on electron doping as in the case of the electron-doped square-lattice cuprate superconductors. It has been argued that the triangular-lattice cobaltate superconductors are probably the only system other than the square-lattice cuprate superconductors where a doped Mott insulator becomes a superconductor \cite{Baskaran03,Wang04,Liu04}. Furthermore, the triangular-lattice cobaltate superconductors also introduce a frustration of the interaction between electrons, and therefore an interplay between superconductivity and geometrical frustration is invoked. In this case, some questions are raised: (a) does superconductivity appear with the hole doping? (b) is the maximum SC transition temperature in the electron-doped case also lower than that in the hole-doped side? (c) does this geometric frustration suppress $T_{\rm c}$ to low temperatures?

To see within a unified physical picture whether the behaviors occurred in the electron-doped triangular-lattice superconductors also emerge in the hole-doped counterparts or not is important, since it refers to whether the nature of the particle-hole asymmetry in a doped Mott insulator is universal or not. In this paper, we try to study this issue in the doped triangular-lattice Mott insulators based on the kinetic-energy driven SC mechanism. We show that (a) as in the doped square-lattice Mott insulators \cite{Bednorz86,Tokura89,Horio16,Song17}, superconductivity should also appear with the hole doping; (b) however, the maximum SC transition temperature in the electron-doped case is lower than that in the hole-doped side; (c) the geometric frustration, accompanied by large fluctuations, suppresses $T_{\rm c}$ of the triangular-lattice superconductors to low temperatures.

Following the discovery of the square-lattice cuprate superconductors \cite{Bednorz86,Tokura89,Horio16,Song17}, there were various suggestions of way in which the $t$-$J$ model can capture the essential physics of a doped Mott insulator \cite{Anderson87,Phillips10}. The original $t$-$J$ model consists of two parts \cite{Anderson87}, the kinetic energy part includes the nearest-neighbor (NN) hopping term $t$, while the magnetic energy part is described by a Heisenberg term with the NN spin-spin antiferromagnetic (AF) exchange $J$. However, the electronic properties and electron-hole asymmetry in a doped Mott insulator on a triangular lattice may be better accounted for by including the next NN hoping $t'$. In this case, we start from the $t$-$t'$-$J$ model on a triangular lattice,
\begin{eqnarray}\label{tjmodel}
H&=&-t\sum_{l\hat{\eta}\sigma}C_{l\sigma}^{\dagger}C_{l+\hat{\eta}\sigma}+t'\sum_{l\hat{\tau}\sigma}C_{l\sigma}^{\dagger}C_{l+\hat{\tau}\sigma}\nonumber\\
&&+\mu \sum_{l\sigma}C_{l\sigma }^{\dagger}C_{l\sigma }+J\sum_{l\hat{\eta}}{\bf S}_{l}\cdot{\bf S}_{l+\hat{\eta}},
\end{eqnarray}
where the summation is over all sites $l$, and for each $l$, over its NN sites $\hat{\eta}$ or next NN sites $\hat{\tau}$, $C^{\dagger}_{l\sigma}$ ($C_{l\sigma}$) is electron creation (annihilation) operator with spin $\sigma$, ${\bf S}_{l}=(S^{x}_{l},S^{y}_{l}, S^{z}_{l})$ are spin operators, and $\mu$ is the chemical potential. For the electron-doped case \cite{Baskaran03,Wang04,Liu04}, we can work in the hole representation via a particle-hole transformation $C_{l\sigma}\rightarrow C^{\dagger}_{l-\sigma}$, and then the difference between the hole- and electron-doped cases can be expressed as the sign difference of the hopping parameters $t$ and $t'$, i.e., $t>0$ and $t'>0$ for the hole-doped case, while $t<0$ and $t'<0$ in the electron-doped side. In this case, the $t$-$t'$-$J$ model (\ref{tjmodel}) in both the hole- and electron-doped cases is always subjected to an important on-site local constraint $\sum_{\sigma}C^{\dagger}_{l\sigma}C_{l\sigma}\leq 1$ to avoid the double occupancy. It has been shown that this single occupancy local constraint can be treated properly within the fermion-spin theory \cite{Feng0494,Feng15}, where the constrained operators $C_{l\sigma}$ can be decoupled as $C_{l\uparrow}= a^{\dagger}_{l\uparrow}S^{-}_{l}$ and $C_{l\downarrow}=a^{\dagger}_{l\downarrow}S^{+}_{l}$, with the spinful fermion operator $a_{l\sigma}= e^{-i\Phi_{l\sigma}}a_{l}$ that keeps track of the charge degree of freedom together with some effects of spin configuration rearrangements due to the presence of the doped charge carrier itself, while the spin operator $S_{l}$ represents the spin degree of freedom, and then the local constraint of no double occupancy is satisfied in the actual calculations. In this fermion-spin representation, the $t$-$t'$-$J$ model (\ref{tjmodel}) can be rewritten as,
\begin{eqnarray}\label{cssham}
H&=&t\sum_{l\hat{\eta}}(a^{\dagger}_{l+\hat{\eta}\uparrow}a_{l\uparrow}S^{+}_{l}S^{-}_{l+\hat{\eta}}+a^{\dagger}_{l+\hat{\eta}\downarrow} a_{l\downarrow} S^{-}_{l}S^{+}_{l+\hat{\eta}})\nonumber\\
&&-t'\sum_{l\hat{\tau}}(a^{\dagger}_{l+\hat{\tau}\uparrow}a_{l\uparrow}S^{+}_{l}S^{-}_{l+\hat{\tau}}
+a^{\dagger}_{l+\hat{\tau}\downarrow} a_{l\downarrow}S^{-}_{l}S^{+}_{l+\hat{\tau}})\nonumber\\
&&-\mu\sum_{l\sigma}a^{\dagger}_{l\sigma}a_{l\sigma}+J_{\rm{eff}}\sum_{l\hat{\eta}}{\bf{S}}_{l}\cdot {\bf{S}}_{l+\hat{\eta}},
\end{eqnarray}
with $J_{\rm eff}=(1-\delta)^{2}J$, and the charge-carrier doping concentration $\delta=\langle a^{\dagger}_{l\sigma}a_{l\sigma}\rangle=\langle a^{\dagger}_{l}a_{l}\rangle$. At half-filling, the $t$-$t'$-$J$ model (\ref{cssham}) is reduced to the AF Heisenberg model on a triangular lattice. In the early discussions, the strong geometry frustration in the AF Heisenberg model on a triangular lattice was expected to completely destroy AF long-range order (AFLRO) and lead to a spin-liquid \cite{Anderson73}. Later, some studies \cite{Zhitomirsky13,Jolicoeur89,Bernu94,Caprioti99}, however, support the appearance of AFLRO. On the other hand, for the case in the AF Heisenberg model on a square-lattice, AFLRO is destroyed by the hole doping with $\delta\sim 0.05-0.07$ for $t/J\sim 2.5 -5$ \cite{Lee88,Khaliullin93}. However, the triangular lattice introduces a strong geometry frustration for the coupling between spins, which could be even more favorable than in the square lattice to formation of the spin-liquid state for sufficiently low doping, and then there is no AFLRO in the doped regime where superconductivity appears.

For an understanding of the physical properties of the square-lattice cuprate superconductors in the SC-state, the kinetic-energy driven SC mechanism has been developed \cite{Feng15,Feng0306,Feng12} based on the $t$-$J$ model in the fermion-spin representation, where in the doped regime without AFLRO, the interaction between charge carriers and spins from the kinetic-energy term in the $t$-$J$ model induces a charge-carrier pairing state (then the electron pairing state) with the d-wave symmetry by the exchange of spin excitations. This SC-state is controlled by both the charge-carrier pair gap and the quasiparticle coherence, which leads to that the maximal $T_{\rm c}$ occurs around the optimal doping, and then decreases in both the underdoped and overdoped regimes. In particular, within this kinetic-energy driven SC mechanism, superconductivity in the electron-doped Mott insulator on a triangular lattice has been discussed based on the $t$-$J$ model in the fermion-spin representation \cite{Liu05,Qin15}, where the obtained optimal $T_{\rm c}$ occurs in a narrow range of the optimal doping, and then decreases for both underdoped and overdoped regimes, in qualitative agreement with the corresponding experimental result \cite{Schaak03} of the electron-doping triangular-lattice cobaltate superconductors Na$_{x}$CoO$_{2}$$\cdot y$H$_{2}$O. Moreover, based on this theoretical framework \cite{Liu05,Qin15}, the electromagnetic response \cite{Qin15}, the thermodynamic properties \cite{Ma16}, and the nature of the electron Fermi surface of the electron-doped Mott insulator on a triangular lattice \cite{Ma161} have been discussed, and then the experimental results of the superfluid density, the specific-heat, and the doping dependence of the electron Fermi surface in the triangular-lattice cobaltate superconductors are qualitatively reproduced. Following these previous discussions \cite{Liu05,Qin15}, the full charge-carrier diagonal and off-diagonal Green's functions of the $t$-$t'$-$J$ model on a triangular-lattice (\ref{cssham}) in the SC-state can be obtained as,
\begin{widetext}
\begin{subequations}\label{Green-functions}
\begin{eqnarray}
g({\bf k},\omega)&=&{1\over\omega-\xi_{\bf k}-\Sigma^{({\rm a})}_{1}({\bf k},\omega)-[\Sigma^{({\rm a})*}_{2}({\bf k},\omega)\Sigma^{({\rm a})}_{2} ({\bf k},\omega)]/[\omega+\xi_{\bf k}+\Sigma^{({\rm a})}_{1}({\bf k},-\omega)]},\label{DEGF}\\
\Gamma^{\dagger}({\bf k},\omega)&=&-{\Sigma^{({\rm a})}_{2}({\bf k},\omega)\over [\omega-\xi_{\bf k}-\Sigma^{({\rm a})}_{1}({\bf k},\omega)] [\omega+ \xi_{\bf k}+\Sigma^{({\rm a})}_{1}({\bf k},-\omega)]-\Sigma^{({\rm a})*}_{2}({\bf k},\omega)\Sigma^{({\rm a})}_{2}({\bf k},\omega)},\label{ODEGF}
\end{eqnarray}
\end{subequations}
\end{widetext}
where the mean-field (MF) charge-carrier excitation spectrum $\xi_{\bf k}=Zt\chi_{1} \gamma_{\bf k}-Zt'\chi_{2} \gamma_{\bf k}'-\mu$, with the number of the NN or next NN sites $Z$, $\gamma_{\bf k}=[\cos k_{x}+2\cos(k_{x}/2)\cos (\sqrt{3} k_{y}/2)]/3$, $\gamma_{\bf k}'=[\cos (\sqrt{3} k_{y})+ 2\cos(3k_{x}/2)\cos (\sqrt{3} k_{y}/2)]/3$, the spin correlation functions $\chi_{1}=\langle S^{+}_{l}S^{-}_{l+\hat{\eta}}\rangle$ and $\chi_{2}=\langle S^{+}_{l}S^{-}_{l+\hat{\tau}}\rangle$.

In the strong coupling formalism \cite{Eliashberg60,Mahan81}, the charge-carrier self-energy $\Sigma^{({\rm a})}_{2}({\bf k},\omega)$ in the particle-particle channel represents the charge-carrier pair gap $\bar{\Delta}^{({\rm a})}({\bf k})=\Sigma_{2}({\bf k},\omega=0)$. Experimentally, the observed specific-heat data \cite{Oeschler08} are consistent with these fitted results obtained from phenomenological Bardeen-Cooper-Schrieffer formalism with the d-wave ($d_{1{\bf k}}+ id_{2{\bf k}}$) symmetry without gap nodes, where $d_{1{\bf k}}=2{\cos}k_{x}-{\cos}[(k_{x}-\sqrt{3}k_{y}) /2]-{\cos}[(k_{x}+\sqrt{3}k_{y}) /2]$ and $d_{2{\bf k}}=\sqrt{3}{\cos}[(k_{x}+\sqrt{3}k_{y})/2]-\sqrt{3}{\cos}[(k_{x}-\sqrt{3} k_{y})/2]$. Theoretically, the numerical simulations \cite{Lee90,Honerkamp03,Watanabe04,Weber06,Zhou08,Kiesel13} indicates that the lowest energy state of the AF Heisenberg model on triangular lattice is the d-wave $(d_{1{\bf k}}+id_{2{\bf k}})$ state with the energy gap $\Delta_{\bf k}\propto\Delta(d_{1{\bf k} } + id_{2{\bf k}})$. In this case, the SC-state symmetry can be chosen as the d-wave pairing symmetry $\bar{\Delta}^{({\rm a})}_{\bf k}= \bar{\Delta}^{({\rm a})}(d_{1{\bf k}}+id_{2{\bf k}})$ as in the previous discussions \cite{Qin15}. On the other hand, the charge-carrier self-energy $\Sigma^{({\rm a})}_{1}({\bf k},\omega)$ in the particle-hole channel can be separated into two parts: $\Sigma^{({\rm a})}_{1}({\bf k},\omega)= \Sigma^{({\rm a} )}_{\rm 1e}({\bf k},\omega)+\omega\Sigma^{({\rm a})}_{\rm 1o}({\bf k},\omega)$, with $\Sigma^{({\rm a})}_{\rm 1e}({\bf k},\omega)$ and $\Sigma^{({\rm a})}_{\rm 1o}({\bf k},\omega)$ that are, respectively, the corresponding symmetric and antisymmetric parts. In particular, the antisymmetric part $\Sigma^{({\rm a})}_{\rm 1o}({\bf k},\omega)$ is directly related to the charge-carrier quasiparticle coherent weight \cite{Qin15} as $Z^{-1}_{\rm aF}=1-{\rm Re} \Sigma^{({\rm a})}_{\rm 1o}({\bf k},\omega=0)\mid_{{\bf k}_{0}}$, with ${\bf k}_{0}=[\pi/3,\sqrt{3}\pi/3]$. In this case, the electron self-energies $\Sigma^{({\rm a})}_{1}({\bf k},\omega)$ in the particle-hole channel and $\Sigma^{({\rm a})}_{2}({\bf k},\omega)$ in the particle-particle channel can be evaluated in terms of the spin bubble as \cite{Qin15},
\begin{widetext}
\begin{subequations}\label{SE1}
\begin{eqnarray}
\Sigma^{({\rm a})}_{1}({\bf k},\omega)&=&{1\over N^{2}}\sum_{{\bf pp'}n}(-1)^{n+1}\Omega^{({\rm a})}_{\bf pp'k}
\left [U^{2}_{{\rm a}{\bf p}+{\bf k}} \left ({F^{(n)}_{{\rm 1a} {\bf p p'k}}\over\omega+\omega_{n{\bf p}{\bf p}'}-E_{{\rm a}{\bf p}+{\bf k}}}\right .\right .
+\left . {F^{(n)}_{{\rm 2a}{\bf pp'k}} \over\omega- \omega_{n{\bf p}{\bf p}'}-E_{{\rm a}{\bf p}+{\bf k}}}\right )\nonumber\\
&&+V^{2}_{{\rm a}{\bf p}+{\bf k}} \left ({F^{(n)}_{{\rm 1a}{\bf pp'k}}\over\omega-\omega_{n{\bf p}{\bf p}'}+E_{{\rm a}{\bf p}+{\bf k}}} \right .
+\left .\left. {F^{(n)}_{{\rm 2a}{\bf pp'k}} \over \omega+\omega_{n{\bf p}{\bf p}'}+E_{{\rm a}{\bf p}+{\bf k}}}\right )\right ],~~~~~\label{PHSE}\\
\Sigma^{({\rm a})}_{2}({\bf k},\omega)&=&{1\over N^{2}}\sum_{{\bf pp'}n}(-1)^{n}\Omega^{({\rm a})}_{\bf pp'k}{\bar{\Delta}^{({\rm a})}_{{\rm Z}{\bf p}
+{\bf k}}\over 2E_{{\rm a}{\bf p } +{\bf k}}}
\left [\left ({F^{(n)}_{{\rm 1a}{\bf pp'k}}\over\omega+\omega_{n{\bf p}{\bf p}'}-E_{{\rm a}{\bf p} +{\bf k}}}\right .\right .
+\left.{F^{(n)}_{{\rm 2a}{\bf pp'k}}\over\omega-\omega_{n{\bf p} {\bf p}'}-E_{{\rm a}{\bf p}+{\bf k}}}\right )\nonumber\\
&&-\left ({F^{(n)}_{{\rm 1a}{\bf pp'k}} \over\omega-\omega_{n{\bf p}{\bf p}'}+E_{{\rm a}{\bf p}+{\bf k}}}\right.
+\left.\left. {F^{(n)}_{{\rm 2a}{\bf pp'k}} \over\omega+\omega_{n{\bf p}{\bf p}'}+ E_{{\rm a}{\bf p}+{\bf k}}}\right )\right ],
\end{eqnarray}
\end{subequations}
\end{widetext}
respectively, where $n=1,2$, $\Omega^{({\rm a})}_{\bf pp'k}=Z_{\rm aF}(Zt\gamma_{{\bf p}+{\bf p}'+{\bf k}}-Zt'\gamma_{{\bf p}+{\bf p}'+{\bf k}}')^{2} B_{{\bf p}'}B_{{\bf p}+{\bf p}'}/(4\omega_{{\bf p}'}\omega_{{\bf p}+ {\bf p}'})$, the charge-carrier quasiparticle coherence factors $U^{2}_{{\rm a} {\bf k}}=(1+\bar{\xi}_{\bf k}/E_{{\rm a}{\bf k}})/2$ and $V^{2}_{{\rm a}{\bf k}}=(1-\bar{\xi}_{\bf k}/E_{{\rm a}{\bf k}})/2$, the charge-carrier quasiparticle energy spectrum $E_{{\rm a}{\bf k}}=\sqrt{\bar{\xi}^{2}_{\bf k}+\mid\bar{\Delta}^{({\rm a})}_{{\rm Z}{\bf k}}\mid^{2}}$, with  $\bar{\xi}_{{\bf k}}=Z_{\rm aF}\xi_{{\bf k}}$, $\bar{\Delta}^{({\rm a})}_{{\rm Z}{\bf k}}=Z_{\rm aF} \bar{\Delta}^{({\rm a}) }_{\bf k}$, $B_{\bf k}= \lambda_{1}[2\chi^{\rm z}_{1}(\epsilon\gamma_{\bf k}-1)+\chi_{1}(\gamma_{\bf k}-\epsilon)]-\lambda_{2}(2\chi^{\rm z}_{2}\gamma_{\bf k}'-\chi_{2})$, $B_{z\bf k}=\epsilon\chi_{1}\lambda_{1}(\gamma_{{\bf k}}-1)-\chi_{2}\lambda_{2} (\gamma_{{\bf k}}'-1)$, $\lambda_{1}=2ZJ_{\rm eff}$, $\lambda_{2}= 4Z\phi_{2}t'$, $\epsilon=1+2t\phi_{1}/J_{\rm eff}$, the charge-carrier's particle-hole parameters $\phi_{1}=\langle a^{\dagger}_{l\sigma} a_{l+\hat{\eta}\sigma} \rangle$, $\phi_{2}=\langle a^{\dagger}_{l\sigma}a_{l+\hat{\tau}\sigma}\rangle$, the spin correlation functions $\chi^{\rm z}_{1}=\langle S_{l}^{\rm z}S_{l+\hat{\eta}}^{\rm z}\rangle$, $\chi^{\rm z}_{2}=\langle S_{l}^{\rm z} S_{l+\hat{\tau}}^{\rm z}\rangle$, $\omega_{n{\bf p}{\bf p}'}=\omega_{{\bf p}+ {\bf p}'}-(-1)^{n} \omega_{\bf p'}$, and the functions, $F^{(n)}_{{\rm 1a}{\bf pp'k}}=n_{\rm F}(E_{{\rm a} {\bf p}+{\bf k}})\{1+n_{\rm B} (\omega_{{\bf p}'+{\bf p}})+n_{\rm B}[(-1)^{n+1}\omega_{\bf p'}]\}+n_{\rm B}(\omega_{{\bf p}'+{\bf p}}) n_{\rm B} [(-1)^{n+1}\omega_{\bf p'}]$ and $F^{(n)}_{{\rm 2a}{\bf pp'k}}=[1-n_{\rm F}(E_{{\rm a}{\bf p}+{\bf k}})]\{1+n_{\rm B}(\omega_{{\bf p}'+{\bf p}})+ n_{\rm B}[(-1)^{n+1}\omega_{\bf p'} ]\}+n_{\rm B}(\omega_{{\bf p}'+{\bf p}}) n_{\rm B}[(-1)^{n+1} \omega_{\bf p'}]$, with $n_{\rm B}(\omega)$ and $n_{\rm F}(\omega)$ that are the boson and fermion distribution functions, respectively, while the MF spin excitation spectra,
\begin{subequations}\label{MFSES}
\begin{eqnarray}
\omega^{2}_{\bf k}&=&\lambda_{1}^{2}\left [{1\over 2}\epsilon\left (A_{1}-{1\over 2}\alpha\chi^{\rm z}_{1}-\alpha\chi_{1}\gamma_{\bf k}\right) (\epsilon-\gamma_{\bf k})\right.\nonumber\\
&&+\left. \left (A_{2}-{1\over 2Z}\alpha\epsilon\chi_{1}-\alpha\epsilon\chi^{\rm z}_{1}\gamma_{\bf k}\right)(1-\epsilon\gamma_{\bf k} )\right]\nonumber\\
&&+\lambda_{2}^{2}\left [\alpha\left (\chi^{\rm z}_{2}\gamma_{\bf k}'-{5\over 2Z}\chi_{2}\right) \gamma_{\bf k}'+{1\over 2}\left (A_{3}-{1\over 3} \alpha\chi^{\rm z}_{2}\right )\right ]\nonumber\\
&&+\lambda_{1}\lambda_{2}\left [\alpha\chi^{\rm z}_{1}(1-\epsilon\gamma_{\bf k})\gamma_{\bf k}'+{1\over 2}\alpha(\chi_{1}\gamma_{\bf k}'- C_{3})(\epsilon-\gamma_{\bf k})\right.\nonumber\\
&&+\left.\alpha\gamma_{\bf k}'(C^{\rm z}_{3}-\epsilon\chi^{\rm z}_{2} \gamma_{\bf k})-{1\over 2}\alpha\epsilon (C_{3}-\chi_{2}\gamma_{\bf k})\right ],\\
\omega^{2}_{z\bf k}&=&\epsilon\lambda^{2}_{1}\left(\epsilon A_{1}-{1\over Z}\alpha\chi_{1}-\alpha\chi_{1}\gamma_{{\bf k}}\right)(1-\gamma_{{\bf k}})\nonumber\\
&&+ \lambda^{2}_{2}A_{3}(1-\gamma_{{\bf k}}')+\lambda_{1}\lambda_{2}\alpha[\epsilon C_{3}(\gamma_{{\bf k}}+\gamma_{{\bf k}}'-2)\nonumber\\
&&+\chi_{2}\gamma_{{\bf k}} (1 -\gamma_{{\bf k}}')],
\end{eqnarray}
\end{subequations}
where $A_{1}=\alpha C_{1}+(1-\alpha)/(2Z)$, $A_{2}=\alpha C^{\rm z}_{1}+(1-\alpha)/(4Z)$, $A_{3}=\alpha C_{2}+(1-\alpha) /(2Z)$, the spin correlation functions $C_{1}=(1/Z^{2})\sum_{\hat{\eta},\hat{\eta'}}\langle S_{l+\hat{\eta}}^{+}S_{l+\hat{\eta'}}^{-}\rangle$, $C^{\rm z}_{1}=(1/Z^{2}) \sum_{\hat{\eta},\hat{\eta'}}\langle S_{l+\hat{\eta}}^{z}S_{l+\hat{\eta'}}^{z}\rangle$, $C_{2}=(1/Z^{2})\sum_{\hat{\tau},\hat{\tau'}}\langle S_{l+\hat{\tau}}^{+}S_{l+\hat{\tau'}}^{-}\rangle$, $C_{3}=(1/Z) \sum_{\hat{\tau}}\langle S_{l+\hat{\eta}}^{+}S_{l+\hat{\tau}}^{-}\rangle$, and $C^{\rm z}_{3}=(1/Z)\sum_{\hat{\tau}}\langle S_{l+\hat{\eta}}^{\rm z}S_{l+\hat{\tau}}^{\rm z}\rangle$. In order not to violate the sum rule of the correlation function $\langle S^{+}_{l} S^{-}_{l}\rangle=1/2$ in the case without an AFLRO, an important decoupling parameter $\alpha$ has been introduced in the decoupling approximation for obtaining the MF spin Green's function, which can be regarded as the vertex correction \cite{Liu05,Qin15}.

In this case, the charge-carrier quasiparticle coherent weight $Z_{\rm aF}$ and charge-carrier pair gap parameter $\bar{\Delta}^{({\rm a})}$ satisfy following two self-consistent equations,
\begin{subequations}\label{SCE1}
\begin{eqnarray}
{1\over Z_{\rm aF}}&=&1+{1\over N^{2}}\sum_{{\bf pp'}n}(-1)^{n+1}\Omega^{({\rm a})}_{{\bf pp'}{\bf k}_{0}}\left [{F^{(n)}_{{\rm 1a}{\bf pp}'{\bf k}_{0}}\over(\omega_{n{\bf p}{\bf p
}' }-E_{{\rm a}{\bf p}+{\bf k}_{0}})^{2}}\right.\nonumber\\
&&\left.+{F^{(n)}_{{\rm 2a}{\bf pp}'{\bf k}_{0}}\over (\omega_{n{\bf p}{\bf p}'}+E_{{\rm a}{\bf p}+{\bf k}_{0}})^{2}}\right ], ~~~~~~~~\label{CCQCWSCE}\\
1&=&{6\over N^{3}}\sum_{{\bf pp'k}n}(-1)^{n}Z_{\rm aF}\Omega^{({\rm a})}_{\bf pp'k}{\Lambda^{({\rm d})*}_{\bf k}\Lambda^{({\rm d})}_{{\bf p}+{\bf k}}\over E_{{\rm a}{\bf p}+{\bf k} }}\nonumber\\
&&\times\left ({F^{(n)}_{{\rm 1a}{\bf pp'k}}\over\omega_{n{\bf p}{\bf p}'}-E_{{\rm a}{\bf p}+{\bf k}}}
-{F^{(n)}_{{\rm 2a}{\bf pp'k}}\over\omega_{n{\bf p}{\bf p}'}+E_{{\rm a}{\bf p}+{\bf k} }}\right ), ~~~~~~~~\label{CCPGPSCE}
\end{eqnarray}
\end{subequations}
respectively, where $\Lambda^{({\rm d})}_{\bf k}=d_{1{\bf k}}+id_{2{\bf k}}$. These two equations (\ref{CCQCWSCE}) and (\ref{CCPGPSCE}) must be solved simultaneously with following self-consistent equations,
\begin{subequations}\label{SCE2}
\begin{eqnarray}
\phi_{1}&=&{1\over 2N}\sum_{\bf k}\gamma_{\bf k}Z_{\rm aF}\left (1-{\bar{\xi}_{\bf k}\over E_{{\rm a}\bf k}}\tanh\left [{1\over 2}\beta E_{{\rm a} \bf k} \right ]\right ),~~~~\\
\phi_{2}&=&{1\over 2N}\sum_{\bf k}\gamma_{\bf k}'Z_{\rm aF}\left (1-{\bar{\xi}_{\bf k}\over E_{{\rm a}\bf k}}\tanh\left [{1\over 2}\beta E_{{\rm a} \bf k}\right ]\right ),\\
\delta&=&{1\over 2N}\sum_{\bf k}Z_{\rm aF}\left (1-{\bar{\xi}_{\bf k}\over E_{{\rm a}\bf k}}\tanh\left [{1\over 2}\beta E_{{\rm a}\bf k}\right ]\right ),\\
\chi_{1}&=&{1\over N}\sum_{\bf k}\gamma_{\bf k}{B_{\bf k}\over 2\omega_{\bf k}}\coth \left [{1\over 2}\beta\omega_{\bf k}\right ],\\
\chi_{2}&=&{1\over N}\sum_{\bf k}\gamma_{\bf k}'{B_{\bf k}\over 2\omega_{\bf k}}\coth \left [{1\over 2}\beta\omega_{\bf k}\right ],\\
C_{1}&=&{1\over N}\sum_{\bf k}\gamma^{2}_{\bf k}{B_{\bf k}\over 2\omega_{\bf k}}\coth \left [{1\over 2}\beta\omega_{\bf k}\right ],\\
C_{2}&=&{1\over N}\sum_{\bf k}\gamma'^{2}_{\bf k}{B_{\bf k}\over 2\omega_{\bf k}}\coth \left [{1\over 2}\beta\omega_{\bf k}\right ],\\
C_{3}&=&{1\over N}\sum_{\bf k}\gamma_{\bf k}\gamma_{\bf k}'{B_{\bf k}\over 2\omega_{\bf k}}\coth \left [{1\over 2}\beta\omega_{\bf k}\right ],\\
{1\over 2}&=&{1\over N}\sum_{\bf k}{B_{\bf k} \over 2\omega_{\bf k}}\coth \left [{1\over 2}\beta\omega_{\bf k}\right ],\\
\chi^{z}_{1}&=&{1\over N}\sum_{\bf k}\gamma_{\bf k}{B_{z\bf k}\over 2\omega_{z\bf k}}\coth \left [{1\over 2}\beta\omega_{z\bf k} \right ],\\
\chi^{z}_{2}&=&{1\over N}\sum_{\bf k}\gamma_{\bf k}'{B_{z\bf k}\over 2\omega_{z\bf k}}\coth \left [{1\over 2}\beta\omega_{z\bf k} \right ],
\end{eqnarray}
\begin{eqnarray}
C^{z}_{1}&=&{1\over N}\sum_{\bf k}\gamma^{2}_{\bf k}{B_{z\bf k}\over 2\omega_{z\bf k}}\coth \left [{1\over 2}\beta\omega_{z\bf k} \right ],\\
C^{z}_{3}&=&{1\over N}\sum_{\bf k}\gamma_{\bf k}\gamma_{\bf k}'{B_{z\bf k}\over 2\omega_{z\bf k}}\coth \left [{1\over 2}\beta\omega_{z\bf k} \right ],
\end{eqnarray}
\end{subequations}
then all the order parameters, the decoupling parameter $\alpha$, and the chemical potential $\mu$ are determined self-consistently without using any adjustable parameters \cite{Liu05,Qin15}.

\begin{figure}[h!]
\centering
\includegraphics[scale=0.4]{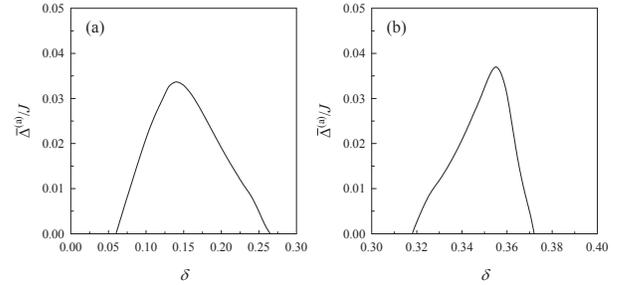}
\caption{The charge-carrier pair gap parameter as a function of doping at $T=0.001J$ for (a) the hole doping with $t/J=3.0$ and $t'/t=0.5$ and (b) the electron doping with $t/J=-3.0$ and $t'/t=0.5$.}
\label{pair-gap-parameter}
\end{figure}

We are now ready to discuss the asymmetric doping dependence of superconductivity between the hole- and electron-doped triangular-lattice superconductors. In Fig. \ref{pair-gap-parameter}, we plot the charge-carrier pair gap parameter $\bar{\Delta}^{(\rm a)}$ as a function of doping at temperature $T=0.001J$ for (a) the hole doping with parameters $t/J=3.0$ and $t'/t=0.5$ and (b) the electron doping with parameters $t/J=-3.0$ and $t'/t=0.5$. It is shown clearly that the charge-carrier pair gap (then SC gap) has a dome-like shape doping dependence in both the hole- and electron-doped cases. In particular, the charge-carrier pair state exists over a broad hole doping range from the hole doping concentrations $\delta\approx0.06$ to $\delta\approx0.26$, where the maximal value of the charge-carrier pair gap appears around the optimal doping $\delta\approx0.14$. However, the charge-carrier pair state emerges only in a narrow electron doping range from the electron doping concentrations $\delta\approx0.32$ to $\delta\approx0.37$, where the maximal value of the charge-carrier pair gap appears around the optimal doping $\delta\approx0.35$. These results are in analogy to the corresponding case of the hole- and electron-doped square-lattice superconductors \cite{Bednorz86,Tokura89,Horio16,Song17}.

\begin{figure}[h!]
\centering
\includegraphics[scale=0.4]{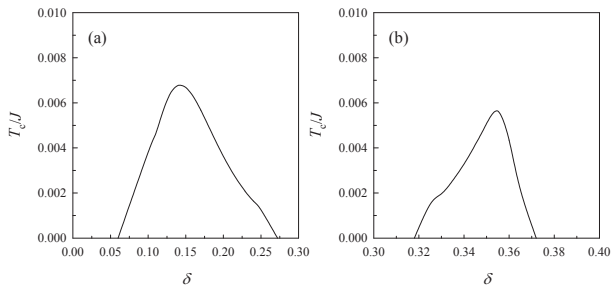}
\caption{The superconducting transition temperature as a function of doping for (a) the hole doping with parameters $t/J=3.0$ and $t'/t=0.5$ and (b) the electron doping with parameters $t/J=-3.0$ and $t'/t=0.5$.}
\label{Tc-doping}
\end{figure}

The SC transition temperature $T_{\rm c}$ on the other hand can be obtained self-consistently from the self-consistent equations (\ref{SCE1}) and (\ref{SCE2}) by the condition $\bar{\Delta}^{\rm (a)}=0$, and the result of $T_{\rm c}$ as a function of doping is plotted in Fig. \ref{Tc-doping} for (a) the hole doping with $t/J=3.0$ and $t'/t=0.5$ and (b) the electron doping with $t/J=-3.0$ and $t'/t=0.5$. Obviously, the doping dependence of $T_{\rm c}$ is consistent with the doping dependence of the charge-carrier pair gap parameter $\bar{\Delta}^{(\rm a)}$ in both the hole- and electron-doped cases shown in Fig. \ref{pair-gap-parameter}. $T_{\rm c}$ arrives at its maximal value around the optimal doping $\delta=0.14$ in the hole-doped case, while it exhibits a maximal value around $\delta\approx0.35$ in the electron-doped side. In particular, the charge-carrier pair state (then the SC-state) in both the hole- and electron-doped triangular-lattice superconductors is controlled by both the charge-carrier pair gap and quasiparticle coherence as shown in the self-consistent equations in Eq. (\ref{SCE1}), which therefore leads to that the maximal $T_{\rm c}$ occurs around the {\it optimal doping}, and then decreases in both the underdoped and the overdoped regimes. However, the maximal $T_{\rm c}$ in the electron-doped case is smaller than that in the hole-doped side. Moreover, the doping range for $T_{\rm c}$ in the electron-doped case is narrower than that in the hole-doped side, reflecting the existence of an asymmetric doping dependence of superconductivity between the hole- and electron-doped triangular-lattice superconductors. In comparison with the corresponding results of the hole- and electron-doped square-lattice cuprate superconductors \cite{Bednorz86,Tokura89,Horio16,Song17}, it is therefore shown that although the dome-shaped doping dependence of $T_{\rm c}$ in the hole- and electron-doped triangular-lattice superconductors is similar to that of the corresponding hole- and electron-doped square-lattice cuprate superconductors, respectively, the $T_{\rm c}$ in both the hole- and electron-doped triangular-lattice superconductors is strongly suppressed by the geometric frustration to much low temperatures, reflecting a fact that the geometric frustration antagonizes superconductivity. This is also why $T_{\rm c}$ in the electron-doped triangular-lattice cobaltate superconductors is much lower than that in the electron-doped square-lattice cuprate superconductors. The present result of the asymmetric doping dependence of superconductivity in the triangular-lattice superconductors is similar to that in the square-lattice superconductors except for the lower $T_{\rm c}$, indicating that the asymmetric doping dependence of superconductivity between the hole- and electron-doped cases may be a common feature for a doped Mott insulator.

In conclusion, within the framework of the kinetic-energy-driven superconductivity, we have discussed the asymmetric doping dependence of superconductivity between the hole- and electron-doped triangular-lattice superconductors. Our results show that for both the hole- and electron-doped triangular-lattice superconductors, the doping evolution of the SC transition temperature $T_{\rm c}$ exhibits a dome-shaped doping dependence, where $T_{\rm c}$ reaches its maximal value around the optimal doping and then decreases in both the underdoped and overdoped regimes. However, the maximum $T_{\rm c}$ in the optimal doping in the electron-doped case is lower than that of the hole-doped side. Moreover, superconductivity appears over a wide doping range from the doping concentrations $\delta\approx0.06$ to $\delta\approx0.26$ in the hole-doped case, however, it only exists in a narrow range of the doping from the doping concentrations $\delta\approx0.32$ to $\delta\approx0.37$ in the electron-doped side. Incorporating the present result with that obtained in the doped square-lattice Mott insulators \cite{Feng12,Mou18}, it is thus shown that the asymmetric doping dependence of superconductivity between the hole- and electron-doped cases may be a common feature for a doped Mott insulator.

\section*{Acknowledgments}

YL was supported by China Scholarship Council, the Natural Science Foundation of Hunan Province under Grant No. 2015JJ3027, and the Science Foundation of Hengyang Normal University under Grant No. 13B44, LQ was supported by the National Natural Science Foundation of China (NSFC) under Grant No. 11647095, and the Natural Science Foundation from Department of Education at Sichuan Province under Grant No. 17ZB0070, and YW and SF was supported by the National Key Research and Development Program of China under Grant No. 2016YFA0300304, and NSFC under Grant Nos. 11574032 and 11734002.

\end{document}